\documentstyle[sprocl]{article}

\input{psfig.sty}

\def\Journal#1#2#3#4{{#1} {\bf #2}, #3 (#4)}

\def\PRL{\em Phys. Rev. Lett.}
\def\PRD{{\em Phys. Rev.} D}

\def\be{\begin{equation}}
\def\ee{\end{equation}}
\def\bea{\begin{eqnarray}}
\def\eea{\end{eqnarray}}
\def\mpl{{{M_{{\rm Pl}}}}}
\def\eqr#1{{Eq.\ (\ref{#1})}}

\begin{document}

\vspace{5mm}
\title{SUPERHEAVY DARK MATTER~\footnote{Talk presented at PASCOS-98,
Northeastern University, Boston, MA, March, 1998.}}

\author{DANIEL J. H. CHUNG}

\address{ Department of Physics and Enrico Fermi Institute\\
The University of Chicago, Chicago, IL~~60637, USA, and\\
NASA/Fermilab Astrophysics Center\\
Fermi National Accelerator Laboratory, Batavia, IL~~60510, USA\\
E-mail: djchung@theory.uchicago.edu}


\maketitle\abstracts{ 
If there exist fields of mass of the order of $10^{13}$ GeV and large
field inflation occurs, their interaction with classical gravitation
will generate enough particles to give the universe critical density
today {\em regardless} of their nongravitational coupling.
}

\section{Introduction}


In the standard dark matter scenarios, WIMPs are usually considered to
have once been in local thermodynamic equilibrium (LTE), and their
present abundance is determined by their self-annihilation cross
section.  In that case, unitarity and the lower bound on the age of
the universe constrains the mass of the relic to be less than 500
TeV.~\cite{griestkam} On the other hand, if the DM particles never
attained LTE in the past, self-annihilation cross section does not
determine their abundance.  For example, axions, which may never have
been in LTE, can have their abundance determined by the dynamics of
the phase transition associated with the breaking of $U(1)_{PQ}$.

These nonthermal relics (ones that never obtained LTE) are typically
light.  However, there are mechanisms that can produce superheavy
(many orders of magnitude greater than the weak scale) nonthermal
relics.  Some of this is reviewed in Ref.~\cite{ckrnontherm}.  Although
not known at the time when this talk was given, it is now known that
if the DM fields are coupled to the inflaton field, then the mass of
the DM particles that can be naturally produced in significant
abundance after inflation can be as large as $10^{-3} \mpl$ (paper in
preparation).

In this article, I discuss the gravitational production
mechanism~\cite{ckrheavy} which is a generic consequence of any
large field inflationary phase ending.  As Ref.~\cite{ckrheavy} shows,
the nonadiabatic change in the way that the spacetime expands at the
end of any large field inflationary model induces superheavy particle
production gravitationally with sufficient efficiency as to render
those superheavy DM to be a significant component of the energy
density in the universe today.  To turn this around, if stable
superheavy WIMPs within the mass range $0.04-2
\times 10^{-6} \mpl$ exist in the mass spectrum of any particle
physics models, then those particle physics models may be ruled out by
cosmology if the occurrence of large field inflation can be
established.~\footnote{The point is that even if large field inflation
occurred, particles of masses of the order of the inflaton mass can
still pose a threat to consistent cosmology.} Note that this
gravitational production of particles is a generic phenomenon which is
insensitive to the way that the DM particles are coupled, as long as
they are stable.  In Ref.~\cite{ckrheavy}, we also analyze the
particle production's large mass asymptotic dependence which is in
general negligibly dependent on the order of adiabaticity of the
adiabatic boundary conditions used unless there is a discontinuity in
some $n$th derivative of the scale factor for $n \sim {\cal O}(1)$.
It is also important to note that if the dark matter decays with a
lifetime of the order of the age of the universe, they may be
observable through cosmic rays.~\cite{kuzmin} Other observational
consequences for the cases in which the superheavy DM is charged or
strongly interacting are under investigation.\cite{rujula}

\section{Scenario Requirements}

Let me now state the scenario more explicitly and briefly address the
requirements of nonthermalization and stability.  If large field
inflation occurs, then when the universe makes a transition out of the
de Sitter phase, there is a nonadiabatic change in the spacetime
expansion leading to a nonadiabatic change in the frequency of the
Fourier mode defining the particles.  Nonadiabatic change means that
the rate of fractional change in the particle mode frequency becomes
larger than the frequency itself.  This gravitational interaction at
the end of inflation induces mixing between positive and negative
frequency modes, leading to quantum creation of DM particles which we
label by $X$.

Suppose these $X$ particles never attained LTE.  The DM abundance
today can be expressed in terms of the DM abundance $n_X(t_e)$ at the
time $t_e$ of their creation (at the end of inflation) as~\footnote{Of
course, our scenario does not contain any late time second inflation.}
\be
\Omega_X h^2 \approx \Omega_R h^2\:
\left(\frac{T_{RH}}{T_0}\right)\:
\frac{8 \pi}{3} \left(\frac{M_X}{\mpl}\right)\:
\frac{n_X(t_{e})}{\mpl H^2(t_{e})}
\label{eq:omegachi}
\ee
where $H$ is the Hubble velocity, $T_0$ is the temperature today,
$T_{RH}$ is the reheating temperature, and $\Omega_R h^2 \approx 4.31
\times 10^{-5}$ is the fraction of critical energy density that is in
radiation today.  This equation says that for a typical reheating
temperature of $10^9$ GeV, $\Omega_X h^2 \sim 10^{17}
(\rho_X(t_e)/\rho(t_e))$ where $\rho(t_e)$ is the total energy density
and $\rho_X(t_e)$ is the energy density stored in the DM particles.
The fraction of total energy density that needs to be extracted to
saturate the matter density upper bound is indeed very small because
the matter energy density grows with respect to the radiation energy density
as inversely proportional to the temperature, while the temperature
difference between the time of reheating and now is large.  Hence, the
challenge lies in creating a very small density of $X$ particles if these
are to contribute significantly to the DM abundance today.

Once these particles are created at the end of inflation, they must
not reach LTE for this scenario to be distinguishable from the
standard one.  Using \eqr{eq:omegachi} with $\Omega_X h^2 < 1 $ and
estimating a conservative upper bound on the WIMP cross section to be
$M_X^{-2}$, we can estimate
\be
 \frac{n_X \langle \sigma_A |v| \rangle}{H} \leq \frac{ 7 \times
10^{-19}}{ (T_{RH}/10^9 \mbox{GeV})} \frac{(H/\mpl)}{(M_X/\mpl)^3},
\label{eq:thermalize}
\ee
the left hand side of which must be less than one at the end of
inflation to avoid LTE.
Thus, because of the $M_X^{-2}$ suppression of a generic WIMP cross
section, a superheavy particle will decouple in general irrespective
of the exact value of the weak coupling constant.

For the $X$ particles to serve as DM, they must have a lifetime that
is of the order of the age of the universe and be extremely massive.
One possible source of these DM particles is the secluded and the
messenger sectors of gauge mediated SUSY breaking models where large
scale SUSY breaking can give rise to large masses, while at the same
time accidental symmetries analogous to baryon number can give large
lifetime to these particles.~\cite{gaugemed} Other natural
possibilities include theories with discrete gauge
symmetries~\cite{hamaguchi} and string/M theory.~\cite{benakli}

\section{Abundance Calculation Results}

We consider a massive scalar field theory (representing the $X$
particle) interacting only with classical gravitational field having
a homogeneous and isotropic metric of the form $ds^2=a^2(\eta)(
d\eta^2 - d{\bf x}^2)$.  We calculate the particle production for
various toy models corresponding to different functions $a(\eta)$.
Each model possesses varying degrees of nonadiabaticity in the mode
frequency evolution at the point of transition out of the inflationary
phase.  In order to minimize the boundary effects and minimize the
number of particles produced, the toy models are chosen to have the
property of admitting infinite adiabatic order vacua at the in-out
regions, and the $X$ particles are conformally coupled to gravity.
With effectively infinite adiabatic order boundary conditions, we
solve numerically for the quantity $n_X(t_e)$ in
\eqr{eq:omegachi}.  The results are shown in
Fig.~1.
\begin{figure}[t]
\vspace*{-10pt}
\hspace*{35pt} \psfig{figure=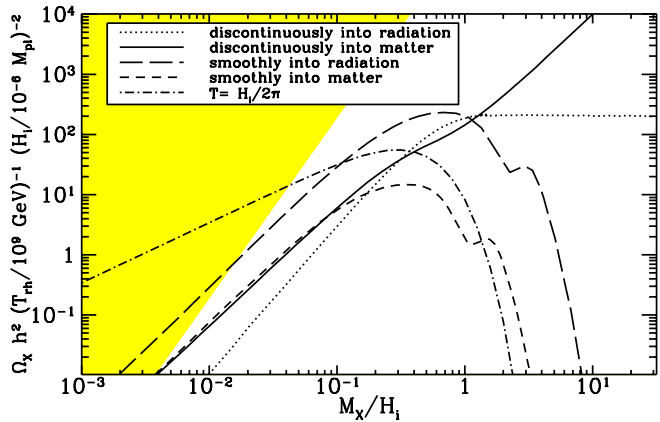,height=2.5in}
\caption{ The dark matter abundance today is shown as a
function of the particle mass for various models.  The mass is given
in terms of $H(\eta_e) \approx 10^{-6} \mpl$ (the Hubble parameter at
conformal time $\eta=\eta_e$, the end of inflation).  In the
``discontinuously into radiation'' case, $a''(\eta)$ has a
discontinuity at $\eta=\eta_e$, while in the ``discontinuously into
matter'' case, $a'(\eta)$ has a discontinuity at $\eta=\eta_e$.  The
curves labeled ``smoothly into'' is for $a(\eta)$ that satisfies
$(d^\nu a/d \eta^\nu)/a^{\nu+1} <
\infty$ for all $\eta$ and natural numbers $\nu$.  The curve labeled
$T=H_i/(2 \pi)$ shows a thermal density with this temperature.  The
unshaded region satisfies the {\it conservative} nonthermalization
condition obtained by considering Eq.\
(\protect{\ref{eq:thermalize}}).
\label{fig:mainresult}}
\end{figure}
The figure clearly shows that for the mass range of $0.04 - 2
\times 10^{-6} \mpl$, $X$ can be naturally produced
gravitationally in cosmologically significant amounts.

\section*{Acknowledgments}
I thank Edward Kolb and Antonio Riotto for their collaboration on the
work presented here.

\end{document}